# Computational and Biological Analogies for Understanding Fine-Tuned Parameters in Physics

Clément Vidal
Center Leo Apostel
Evolution, Complexity and Cognition research group
Vrije Universiteit Brussel (Free University of Brussels)
Krijgskundestraat 33, 1160 Brussels, Belgium
http://clement.vidal.philosophons.com
clement.vidal@philosophons.com

**Abstract:**
In this philosophical paper, we explore computational and biological analogies to address the fine-tuning problem in cosmology. We first clarify what it means for physical constants or initial conditions to be fine-tuned. We review important distinctions such as the *dimensionless* and *dimensional* physical constants, and the classification of constants proposed by Lévy-Leblond. Then we explore how two great analogies, computational and biological, can give new insights into our problem. This paper includes a preliminary study to examine the two analogies. Importantly, analogies are both useful and fundamental cognitive tools, but can also be misused or misinterpreted. The idea that our universe might be modelled as a computational entity is analysed, and we discuss the distinction between physical laws and initial conditions using algorithmic information theory. Smolin introduced the theory of "Cosmological Natural Selection" with a biological analogy in mind. We examine an extension of this analogy involving intelligent life. We discuss if and how this extension could be legitimated.

## Contents

# 0 Introduction[1]

After Leibniz famously wrote "why is there something rather than nothing?" he then qualified this by writing: "Also, given that things have to exist, we must be able to give a reason why they have to exist as they are and not otherwise." (Leibniz 1714, para. 7). Trying nowadays to tackle this age old metaphysical question, we have to take into account the progress that science has made. Both the emerging sciences of complexity and cosmology can help this philosophical enterprise. Modern science can successfully connect the general physico-chemical cosmological evolution with biological and cultural evolution (e.g. Chaisson 2001; De Duve 1995). Thus, it seems reasonable to assume that science is an effective method in enabling us to understand the whole evolution of our universe. The problem of harmony in the cosmos has thus shifted to its beginning : why did the universe start with these initial conditions and laws, and not others?

The belief in God allowed western thinkers to understand why the "Laws of Nature" are as they are and not otherwise. Scientific activity ultimately consisted of discovering the “Laws of Nature” set up by God. However, now that many scientists no longer believe in God, there is a lack of explanation in the origin of the “Laws of Nature” (Davies 1998).

Why is our universe as it is, and not different? This question is a very much debated issue, at the intersection of cosmology, theology and philosophy. In modern terms, it is known as the *fine-tuning problem* in cosmology. It states that *if a number of parameters, both constants in physics and initial parameters in big-bang models had been slightly different, no life or, more generally, no complexity would have emerged* (Barrow and Tipler 1986; Davies 1982; 2008; Ellis 1993; Leslie 1989; 1998; Rees 2000; Hogan 2000; Barrow et al. 2008). The standard models of particle physics and cosmology require 31 free parameters to be specified (Tegmark et al. 2006). It is a main challenge of modern physics to build stronger theories able to reduce this number.

As Leslie (1989) reminds us, the argument that the universe is fine tuned is not based on the assumption that there is a fine-tuner. It *only* means that the emergence of life or complexity is sensitive to many different slight changes in our physical and cosmological models.

The literature around this issue can be divided into two main classes of solutions: “God” or “Multiverse”. Either it is God who created the Universe with all its parameters fit for life and intelligence; or there is a huge number of other universes with different parameters, so that it is very probable that there is one containing life and intelligence. The fact that it is the one we happen to inhabit is an observational selection effect which thus makes fine-tuning less mysterious

1 This paper was commented and criticized by Greben (2009) and Vaas (2009). My response can be found in (Vidal 2009) where this paper also appeared.

(e.g. Carr 2007; Bostrom 2002).

From a rational and scientific point of view, an appeal to God suffers from being a non-naturalistic explanation. Furthermore, God is often linked with the "God of the gaps" assumption. If we can not understand a phenomenon, we use God to explain it, and we thus do not seek another explanation. This attitude can, by its very definition, *explain everything*. A parallel can be made with the seemingly opposite approach to the fine-tuning problem, holding that the universe's parameters happened by pure random chance. Indeed, this appeal to chance works everywhere and is not restricted by any limit; so it can also *explain everything*. One should also note that *in fine*, appealing to a multiverse with a selection effect is similar to the chance explanation. Indeed, in both cases we have the chance to be in a life-permitting universe.

Iris Fry (1995) already pointed out in the context of explaining the origin of life that appealing to chance or to God are after all similar attempts. This situation is comparable with the "God" or "Multiverse" alternative. Both options thus surprisingly suffer from similar shortcomings. Are there other possible roads to address the fine-tuning problem?

This paper aims explicitly and carefully to use two great analogies, to open new research axes. These are *computational* and *biological* analogies. We first clarify what physical constants and initial conditions are, to better grasp what it means to state that they are fine-tuned. We thus review some propositions to classify the two principal sets of physical parameters: physical constants in particle physics and initial conditions in cosmology. Before exploring the two analogies, we investigate the basic functioning of analogical reasoning. We do this to avoid naïve ontological statements such as "the Universe is a computer" or "the Universe is alive". We then analyse aspects of computational and biological analogies. Among others, this leads us to questionning the distinction between laws and initial conditions. We also point out some of the epistemological limits of Smolin's attempt to tackle the fine-tuning problem with his biological-inspired theory of "Cosmological Natural Selection". We then propose a *philosophical* extension to it, "Cosmological *Artificial* Selection" which includes a possible role for intelligent life.

# 1 Physical constants and initial conditions

We distinguish between *dimensional* and *dimensionless* physical constants, as proposed by Michael Duff in (Duff, Okun, and Veneziano 2002). If a constant has a unit after its value, it is dimensional. Dimensional constants depend on our unit system choice and thus have a conventional aspect. The velocity of light $c$, the reduced Planck constant $\hbar$ or the gravitational constant $G$ are dimensional constants (their respective dimensions are, for example, $m.s^{-1}$, eV.s and $m^3.kg^{-1}.s^{-2}$). Certainly, we can for example make the velocity of light equal to 1, and thus apparently dimensionless. However, this applies only to a particular unit system, and the constant will be dimensional again in another unit system.

By contrast, dimensionless constants are dimensionless in *any* unit system. They are ratios between two physical quantities, such as two forces or two masses. For example, the electron-proton mass ratio is $m_e/m_p = 1/1836.15$... Since the two quantities are masses, we can get rid of the units (i.e. the dimension), and keep only a pure number. Other dimensionless constants are deduced by a similar *dimensional analysis*. If the analysis leads to a pure number without dimension, we have a "dimensionless" constant.

Along with this dimensional versus dimensionless distinction, Jean-Marc Lévy-Leblond (1979, 238) proposed another complementary classification of physical constants. Three types are distinguished, in order of increasing generality:

> A. Properties of particular physical objects considered as fundamental constituents of matter; for instance, the masses of "elementary particles", their magnetic moments, etc.
>
> B. Characteristics of classes of physical phenomena: Today, these are essentially the coupling constants of the various fundamental interactions (nuclear, strong and weak, electromagnetic and gravitational), which to our present knowledge, provide a neat partition of all physical phenomena into disjoint classes.
>
> C. Universal constants, that is constants entering universal physical laws, characterizing the most theoretical frameworks, applicable in principle to any physical phenomenon; Planck constant $\hbar$ is a typical example.

The classification is only intended to be a basis for discussing and analysing the historical evolution of different physical constants. For example, the constant *c*, the velocity of light, was first discovered as a type-A constant. It was a property of light, as a physical object. With the work of Kirchhoff, Weber, Kohlrausch and Maxwell, the constant gained type-B status when it was discovered that it also characterized electromagnetic phenomena. Finally, it gained type-C status when special and general relativity were discovered, synthesizing concepts such as spatio-temporal intervals, or mass and energy (see (Lévy-Leblond 1979, 252-258) for a detailed account of the status change of *c*).

What happens next, when a constant has reached its type-C status? The fate of universal constants (type-C), explains Lévy-Leblond, is to "see their nature as concept synthesizers be progressively incorporated into the implicit common background of physical ideas, then to play a role of mere unit conversion factors and often to be finally forgotten altogether by a suitable redefinition of physical units." (Lévy-Leblond 1979, 246). More precisely, this remark leads him to the distinction of three subclasses of type-C constants, according to their historical status:

(i) the *modern* ones, whose conceptual role is still dominant (e.g. $\hbar$, $c$);
(ii) the *classical* ones, whose conceptual role is implicit and which are considered as unit conversion factors (e.g. thermodynamical constants $k$, $J$);
(iii) *archaic* ones, which are so well assimilated as to become invisible (e.g. the now obvious ideas that areas are square of lengths).

If all dimensional constants follow this path, then they all become "archaic", and thus integrated in the background of physical theories. The fate of dimensional constants seems then to fade away. Is it possible to seriously consider this remark, and try to relegate all dimensional constants to archaic ones? Michael Duff (Duff 2002; Duff, Okun, and Veneziano 2002) did a first step in this direction by convincingly arguing that the number of dimensional constants (type-C) can be reduced to ... zero! Thus, he considers constants like $c$, $G$, $\hbar$, which are often considered as "fundamental", as merely unit conversion factors. According to his terminology, only dimensionless constants should be seen as fundamental.

A dimensionless physics approach is also proposed in the framework of scale relativity (Nottale 2003, 16). Following the idea of relativity, one can express any physical expression in terms of ratios. Indeed, in the last analysis a physical quantity is always expressed *relative* to another. Of course, experimentalists still need to refer to metric systems, and often to many more dimensional physical constants than just the common $c$, $G$ and $\hbar$. The point here is that it is possible to express the results in physical equations without reference to those dimensional constants (see also (Lévy-Leblond 1979, 248-251)).

What are the consequences of these insights for fine-tuning arguments? If the fate of dimensional constants is to disappear, then the associated fine-tuning arguments with these constants should also disappear. Considering what would happen if a type-C dimensional constant would be different has to be considered skeptically. Such a scenario has already been considered, for example by Rozental (1980) where he analysed what would happen if the Planck constant were to be increased by 15%. Again, as Duff argued, the problem is that dimensional constants are conventions, and changing them is changing a convention, not physics. It is thus only meaningful to express the possible changes in terms of dimensionless constants.

In fact, most fine-tuning arguments focus on considering changes in dimensionless constants. Typically, they consider what would happen if we were to change one of the four fundamental dimensionless coupling constants. These include $\alpha$ for electromagnetism, $\alpha_G$ for gravity, $\alpha_W$ for the weak nuclear force and $\alpha_s$ for the strong nuclear force. It should be noted that these constants are actually not constant, since they change with energy scale (Wilczek 1999).

The main conclusion in this area of study is that there are *conditions* on these constants for key events in the evolution of complexity in the universe to occur. For example, during the big-bang nucleosynthesis, the condition $\alpha_G < (\alpha_W)^4$ must be fulfilled, or else all hydrogen goes to helium. Bernard Carr (2007a) provided a detailed account of other constraints related to these four constants for the baryosynthesis, nucleosynthesis, star formation and planet formation to occur.

Along with these coupling constants, there is a whole other set of fine-tuning arguments based on cosmological parameters. These include parameters such as matter density (Ω), amplitude of initial density fluctuations, photon-to-baryon ratio, etc. For example, the "total density parameter Ω must lie within an order of magnitude of unity. If it were much larger than unity, the Universe would recollapse on a time-scale much less than the main-sequence time of a star. On the other hand, if it were much smaller than unity, density fluctuations would stop growing before galaxies could bind." (Carr 2007, 80). We wrote in introduction that one of the main challenges of modern physics is to construct theories able to reduce this number of parameters, both in the standard model of particle physics and in cosmological models. Parameters involved in cosmological models can be explained by new physical theories. Such is the case with the dimensionless cosmological constant, whose value has been predicted by scale relativity (Nottale 2008, 27-31).

Following Duff and Lévy-Leblond, we saw that type-C constants are bound to disappear. Another challenge we would like to propose is the following: could type-A and type-B constants emerge from initial conditions in a cosmological model? If we were be able to explain all these constants in terms of a cosmological model, it would certainly be a great achievement. Smolin (1997, 316) also argued that fundamentally, progress in quantum mechanics must lead to a cosmological theory. Indeed, all particles ultimately originate from the big-bang, thus *a complete understanding of particle physics should include an explanation of their origin, and thus relate with a cosmological model.*

In a certain sense, progress in this direction has already happened, if we consider the discovery of big-bang nucleosynthesis. Properties of atomic elements could be thought as fundamental constituents of matter, and thus type-A constants, until we discovered they were actually formed at the big-bang era. If we extrapolate this trend, a future cosmological model may be able to derive many (or even all) type-A constants from initial conditions.

The same can be said about fundamental coupling constants (type-B). Special scale relativity can indeed predict the value of $\alpha_s$ (the strong nuclear force) at the Z mass energy level. This was predicted with great precision, and has been confirmed by experimental measures (see (Nottale 2008, 26-27) in this volume for further details). Thus, if physics continues its progress, it is reasonable to conceive that particle physics models would become integrated into a cosmological model. The consequence for fine-tuning arguments is that *fine-tuning of physical constants would progressively be reduced to initial conditions of a cosmological model.*

Given this analysis of physical constants, let us now examine an argument against the idea of fine-tuning proposed by Chaisson (2006, xvi-xvii):

> Rather than appealing to Providence or "multiverses" to justify the numerical values of some physical constants (such as the speed of light or the charge of an electron), I prefer to reason that when the laws of science become sufficiently robust, we shall naturally understand the apparent "fine-tuning" of Nature. It's much akin to mathematics, when considering the value of π. Who would have thought, a priori, that the ratio of a circle's circumference to its diameter would have the odd value of 3,14159.... ? Why isn't it just 3, or 3,1, or some other crisp number, rather than such a peculiar value that runs on ad infinitum? We now understand enough mathematics to realize that this is simply the way geometry scales; there is nothing mystical about a perfect circle -yet it surely is fine-tuned, and if it were not it wouldn't be a circle. Circles exist as gracefully rounded curves closed upon themselves *because* π has the odd value it does. Likewise, ordered systems in Nature, including life, likely exist *because* the physical constants have their decidedly odd values.

First, we can remark that the speed of light and the charge of the electron are dimensional constants. As we analysed, it makes not much sense to speak about a variation, and thus a fine-tuning of them (see also (Duff 2002)).

Let us look more closely at Chaisson's suggestion that if "the laws of science become sufficiently robust we shall naturally understand the apparent 'fine-tuning' of Nature". Considering what we have argued so far, we can agree with this. We have outlined Duff's proposal that dimensional constants can be reduced to 0. We have suggested that fundamental coupling constants could be in future explained from more general principles, and that many apparent "fundamental constants" in the past can nowadays be explained by more general theories. Accordingly, a great deal of fine-tuning has been and certainly will be explained by more advanced physical theories.

Let us now examine the analogy with π. We can see a disanalogy between mathematical and physical constants. Mathematical constants are defined *a priori* by the axioms: they are *internal to the system* and are generally definable and computable numbers. For example, we have plenty of algorithms to calculate π. This is not the case with physical constants. Many of them remain *external to the system,* in the sense that they are not computable from inside the model. At some stage there has been a measurement process to get their values. We can reformulate Chaisson's position by saying that progress in science will allow us to understand (or compute) these constants, from more fundamental principles. We will develop this computational view in the third section of this paper.

However, if we saw that future physics could understand physical constants in terms of a cosmological model, it is unlikely that this would also include the initial conditions of this model. Indeed, if we were to have a theory deciding all values of initial conditions in a cosmological model, it then leads to the idea of a "final theory" or a "theory of everything". This would bring many conceptual and metaphysical problems (Rescher 2000). One of these problems is that, ironically, this idea of a final theory is an act of faith and is thus similar to the Providence explanation (e.g. Davies 2008, 170). Smolin (1997, 248) wrote that the "belief in a final theory shares with a belief in a god the idea that the ultimate cause of things in this world is something that does not live in the world but has an existence that, somehow, transcends it." Fine-tuning arguments based on initial conditions of a cosmological model thus remain intact in Chaisson's critique.

# 2 Analogies for scientific purposes

We have seen some aspects of physical parameters from a physical point of view. We will shortly turn to other approaches to the fine-tuning issue, inspired by computational and biological analogies. However, let us begin with a short digression explaining how analogies can be used for scientific purposes. Many great scientific discoveries have been triggered by analogies (see (Holyoak and Thagard 1995, chap. 8) for plenty of examples). This constitutes an important motivation to understand in greater detail the functioning of analogical reasoning. Analogical reasoning is certainly an essential cognitive tool, which nevertheless needs to be carefully used.

What is an analogy? It is a structural or functional similarity between two domains of knowledge. For example, a cloud and a sponge are analogous in the sense that they can both hold and give back water. More precisely, we can give the following definition: "an analogy is a mapping of knowledge from one domain (the base) into another (the target) such that a system of relations that holds among the base objects also holds among the target objects." (Gentner and Jeziorski 1993, 448-449). In this very simple example, the relations "holding and giving back water" which are verified in the base (the cloud) are also verified in the target (the sponge).

Analogical reasoning is recognized to be a basic cognitive mechanism allowing us to learn and solve problems (e.g. Minsky 1986; Hofstadter 1995; Holyoak and Thagard 1995). Leary (1990, 2) even argued that language and every kind of knowledge is rooted in metaphorical or analogical thought processes. Indeed, when we do not know a domain at all, we must use analogies as a cognitive tool to potentially gain some insights from what we already know. In this manner, a map from the known to the unknown can be drawn.

Specifically, Holyoak and Thagard (1995, 185-189) argued that analogical reasoning is helpful in *discovering, developing, educating,* or *evaluating* scientific theories. Indeed, they allow us to propose new hypotheses, and thus *discover* new phenomena. These new hypotheses trigger us to *develop* new experiments and theories. Let us note however that there is nothing automatic or easy in this process. The relational system should first be examined in both domains, and then a more precise analogy or disanalogy can be found worthy of testing.

The *educating* part of analogies is useful for diffusing scientific knowledge, both to colleagues and pupils. Indeed, it can be very helpful and efficient to consciously use analogies to help others grasp a new idea, based on what they already know.

The *evaluating* part confronts us with one of the main dangers of analogies. One should emphasize that *an analogy is not a proof*. Analogies can thus not properly be used to prove statements, but their main utility is in giving *heuristics* for discovering and developing scientific theories. To illustrate this point, let us consider the teleological argument of God's existence popularized by William Paley (1802), based on the analogy of our Universe with a watch. It goes as follows:

(1) A watch is a fine-tuned object.
(2) A watch has been designed by a watchmaker.
(3) The Universe is fine-tuned.
(4) The Universe has been designed by God.

In the base domain (1-2), we have two objects, the watch and the watchmaker. They are linked by a "designed by" relationship. In the target domain (3-4), the Universe is like a watch, and God, like a watchmaker. That the relation (1)-(2) is a verifiable fact does not imply at all that the same relation "designed by" in (3)-(4) should be true. There is no causal relationship between the couple (1)-(2) and (3)-(4). This reasoning at most gives us an *heuristic* invitation to ponder whether the universe is fine-tuned. Although it is a logically flawed argument, one can appreciate that this reasoning induces a strong intuitive appeal.

There are in fact other pitfalls associated with analogical reasoning. To avoid them, Gentner and Jeziorski (1993, 450) proposed six principles of analogical reasoning (Gentner 1993, 450):

> 1. **Structural consistency**. Objects are placed in one-to-one correspondence and parallel connectivity in predicates is maintained.
> 2. **Relational focus**. Relational systems are preserved and object descriptions disregarded.
> 3. **Systematicity**. Among various relational interpretations, the one with the greatest depth - that is, the greatest degree of common higher-order relational structure - is preferred.
> 4. **No extraneous associations**. Only commonalities strengthen an analogy. Further relations and associations between the base and target - for example, thematic connections - do not contribute to the analogy.
> 5. **No mixed analogies**. The relational network to be mapped should be entirely contained within one base domain. When two bases are used, they should each convey a coherent system.
> 6. **Analogy is not causation**. That two phenomena are analogous does not imply that one causes the other.

Is it possible to further generalize the use of analogical reasoning into a science which would focus only on the structural or functional aspects of systems? Studying different models in different disciplines having structural or functional similarities leads to the development of very general interdisciplinary scientific frameworks, like the network or systems theory paradigms. Indeed, Ludwig van Bertalanffy defined general systems theory as an interdisciplinary doctrine

“elaborating principles and models that apply to systems in general, irrespective of their particular kind, elements, and 'forces' involved” (Bertalanffy, quoted in (Laszlo 1972, xvii)). In a similar fashion, the study of networks is independent of the nodes and types of relations considered.

To conclude this section, one can use Hesse's (1966, 8) pragmatically valuable distinction between *positive*, *negative* and *neutral* analogies. The positive analogy addresses the question: *what is analogous?* and constitutes the set of relations which hold in the two domains. The negative analogy addresses the question: *what is disanalogous?* and constitutes the set of relations which do not hold in the two domains. Finally, neutral analogies trigger the question: *are the two domains analogous?* To answer this last question, one has to examine or test if such or such relation holds in the target domain.

Given this analysis, we can now carefully explore the fine-tuning issue, aided by computational and biological analogies. Since we have claimed in the first section that fine-tuning arguments could be in future reduced to initial conditions, we will now focus on this aspect.

# 3 The computational universe

The idea that our universe is analogous to a computer is a popular one. We can see it as the modern version of a mechanistic worldview, looking at the universe as a machine. There are various ways to consider this analogy, with cellular automata, e.g. (Wolfram 2002) with quantum computing e.g. (Lloyd 2005), etc. The analogy has been pushed so far that a modern version of idealism has even be considered, i.e. that our universe would actually be run by a computer, and we might be living in a computer simulation (e.g. Bostrom 2003; Martin 2006).

We saw that fine-tuning arguments might ultimately be reduced to initial conditions of a cosmological model. Here, we examine what the initial conditions are from a computational perspective, and discuss the relation between laws and initial conditions. This is conducted within the framework of Algorithmic Information Theory (AIT, (Chaitin 1974; 1987)). We will then argue that computer simulations provide an indispensable tool if we wish to tackle the fine-tuning problem scientifically. We will conclude by pointing out some limitations of this computational analogy.

AIT studies complexity measures on strings. The complexity measure -the Kolmogorov complexity[2]- of an object such as a piece of text is a measure of the computational resources needed to specify the object. Below is a simple example originally presented in the Wikipedia encyclopaedia (2008) :

2 also known as program-size complexity, Kolmogorov-complexity, descriptive complexity, stochastic complexity, or algorithmic entropy.

> consider the following two strings of length 64, each containing only lower-case letters, numbers, and spaces:
>
> abababababababababababababababababababababababababababababababab
> 4c1j5b2p0cv4w1 8rx2y39umgw5q85s7ur qbjfdppa0q7nieieqe9noc4cvafzf
>
> The first string admits a short English language description, namely "ab 32 times", which consists of 11 characters. The second one has no obvious simple description (using the same character set) other than writing down the string itself, which has 64 characters.

In this AIT framework, laws represent information which can be greatly shortened by algorithmic compression (like the “ab 32 times” string); whereas initial conditions represent information which cannot be so compressed (like the second string). If we import this analogy into physics, a physical law is to be likened to a simple program able to give a compressed description of some aspects of the world; whereas initial conditions are data that we do not know how to compress.

Can we interpret this distinction between physical laws and initial conditions in a cognitive manner? We either express our knowledge in terms of laws if we can compress information, and in terms of initial conditions if we cannot. In this view, scientific progress allows us to dissolve initial conditions into new theories, by using more general and efficient algorithmic compression rules.

In mathematics, Gödel's limitation theorems state that in any sufficiently rich logical system, there will remain undecidable propositions *in that system*. But using another stronger system, one can decide such previously "undecidable" propositions (even if new undecidable propositions will arise in the stronger system...). For example, the consistency of Peano's arithmetic cannot be shown to be consistent within arithmetic, but can be shown to be consistent *relative* to modern set theory (ZFC).

There is a theorem similar to Gödel's incompleteness in AIT. Informally, it states that a computational system cannot compress structure in a system that is more algorithmically complex than this computational system. Let us assume again that physical laws represent compressible information, and initial conditions incompressible information. Are initial conditions in cosmological models algorithmically incompressible? There are two ways to answer this question.

First, we can interpret this incompressible data in an absolute way. This data is then “lawless, unstructured, patternless, not amenable to scientific study, incompressible” (Chaitin 2006, 64). Suggesting that those initial conditions are incompressible implicitly implies that we, poor humans, will never be able to understand them. This attitude freezes scientific endeavour and thus has to be rejected. Limitation theorems are only valid within formal systems, because one needs the system to be completely formalized and specific formal tools to be able to prove them. Therefore, we should be extremely careful when exporting limitation theorems into other less formalized domains. Moreover, the history of science has shown that it is hazardous to fix boundaries on human understanding. Let us take the example of infinity, which was for many centuries thought to be understandable only by a God who is infinite, and not by finite humans. A rigorous theory of infinite numbers, constituting the foundations of modern mathematics, has finally been proposed by the mathematician Georg Cantor. Therefore, boundaries are likely to be broken. We will shortly see how the *multiverse hypothesis* or computer *universe simulation* bring us beyond the apparently incompressible initial conditions.

The second option is that incompressible information may reflect the limits of our theoretical models. If we are not able to account for the reasons of initial conditions, it is a hint that we need a broader theoretical framework to understand them. This situation can be illustrated by considering the problem of the origin of life. In this context, initial conditions for life to emerge are generally assumed without justification: chemical elements are assumed to be here, along with an Earth with water, neither too far nor too near from the Sun, etc. With these hypotheses (and others), we try to explain the origin of life. Now, what if we try to explain the origin of these initial suitable conditions for life? We would then need a broader theory, which in this case is a theory of cosmic evolution. If we then aim to explain initial conditions in cosmology, we are back to the problem of fine-tuning.

As we wrote in the introduction, multiverse models are precisely attempting to introduce a broader theory to explain those initial conditions, by proposing the existence of various other possible universes with different initial conditions. The problem is that the multiverse hypothesis is a *metaphysical assumption*. George Ellis (2007a, 400) expressed it well:

> There can be no direct evidence for the existence of other universes in a true multiverse, as there is no possibility of even an indirect causal connection. The universes are completely disjoint and nothing that happens in one can affect what happens in another. Since there can be no direct or indirect evidence for such systems, what weight does the claim for their existence carry? Experimental or observational testing requires some kind of causal connection between an object and an experimental apparatus, so that some characteristic of the object affects the output of the apparatus. But in a true multiverse, this is not possible. No scientific apparatus in one universe can be affected in any way by any object in another universe. The implication is that the supposed existence of true multiverses can only be a metaphysical assumption. It cannot be a part of science, because science involves experimental or observational tests to enable correction of wrong theories. However, no such tests are possible here because there is no relevant causal link.

To improve testability, Ellis further suggests examining a variation on the causally disconnected universes, considering multi-domain universes that are not causally disconnected. Still, I would like to emphasize the *philosophical* importance of the multiverse hypothesis, because it is a logically consistent way to tackle the fine-tuning problem. Is there a way other than the multiverse to theorize about “other possible universes”? This is what we will analyse now.

One of the main limitations of fine-tuning arguments is that the vast majority of them vary *one single* parameter of physical or cosmological models and conclude that the resulting universe is not fit for developing complexity. What if we would change *several* parameters at the same time? For example, if the expansion rate of the universe had been greater, and gravity had been stronger, could the two changes cancel each other out? Systematically exploring those possibilities seems like a very cumbersome enterprise. As Gribbin and Rees wrote (1991, 269):

> If we modify the value of one of the fundamental constants, something invariably goes wrong, leading to a universe that is inhospitable to life as we know it. When we adjust a second constant in an attempt to fix the problem(s), the result, generally, is to create three new problems for every one that we "solve". The conditions in our universe really do seem to be uniquely suitable for life forms like ourselves, and perhaps even for any form of organic complexity.

A way to overcome this problem would be to use *computer simulations* to test systematical modifications of constants' values. An early attempt of such an approach has been proposed by Victor Stenger (1995; 2000). He has performed a remarkable simulation of possible universes. He considers four fundamental constants, the strength of electromagnetism $\alpha$; the strong nuclear force $\alpha_s$, and the masses of the electron and the proton. He then analysed “100 universes in which the values of the four parameters were generated randomly from a range five orders of magnitude above to five orders of magnitude below their values in our universe, that is, over a total range of ten orders of magnitude” (Stenger 2000). The distribution of stellar lifetimes in those universes shows that most universes have stars that live long enough to allow stellar evolution and heavy elements nucleosynthesis. Anthony Aguirre (2001) reached a similar conclusion by proposing “a class of cosmologies (based on the classical ‘cold big-bang’ model) in which some or all of the cosmological parameters differ by orders of magnitude from the values they assume in the standard hot big-bang cosmology, without precluding in any obvious way the existence of intelligent life.” Alejandro Jenkins and Gilad Perez also recently argued that different universes might be habitable (Jenkins and Perez 2010). In conclusion, other possible universes may also be fine-tuned!

It is certainly possible to critique Stenger's simulation as being too simplistic. Maybe we should consider other or more “fundamental” constants to vary; or to vary the laws of physics themselves. Stenger did not attempt to vary physical laws and it seems indeed a very difficult enterprise, because we do not even know how to make them vary (see Vaas 1998).

In fact, *the distinction between laws and boundary conditions is fuzzy in cosmology* (Ellis 2007b, sec. 7.1). One can see boundary conditions as imposing constraints, not only on initial conditions (lower boundary of the domain), but also at the extremes of the domain. Both physical laws and boundary conditions play the same role of imposing constraints on the system at hand. Because we can not re-run the tape of the universe, it is difficult -if not impossible- to distinguish the two. In this view, some laws of physics might be interpreted as regularities of interactions progressively emerging out of a more chaotic state. The cooling down of the universe would progressively give rise to more stable dynamical systems, which can be described by simple mathematical equations.

A similar situation occurs in computer science. One can distinguish between a program, which is a set of instructions, and the data on which the program operates. The program is analogous to physical laws, and the data to initial conditions. This distinction in computer science can be blurred, when considering self-modifying programs, i.e. a program which modifies itself. Also, at a lower level, both the program and the data are processed in the form of bits, and here also the distinction is blurred.

Back to Stenger's simulation, it does not answer the following questions: would other interesting complex structures like planetary systems, life, intelligence, etc. evolve in those other universes? However, this is only an early attempt in simulating other possible universes, and the enterprise is certainly worth pursuing. The fine-tuning problem could then be seriously tackled, because we would know precisely the likelihood of having our universe as it is, by comparing it to other possible universes. To summarize, this approach needs to:

(1) define a space of possible universes
(2) vary parameters defining this space, to see how likely it is to obtain a universe fine-tuned to develop complex structures.

Many multiverse scenarios such as those in (Carr 2007) proposed answers to step (1). The second proposed step can be tackled with computer simulations.

I argued in (Vidal 2008) that a simulation of an entire universe will result from future scientific activity. Such a simulation would enable us not only to understand our own universe (with “real-world modelling”, or processes-as-we-know-them) but also other *possible* universes (with “artificial-world modelling”, or processes-as-they-could-be). In this way, we would be able to scientifically assess to what degree our universe is fine-tuned or not. If it turns out that our universe is not so special, then a response to fine-tuning would be a principle of *fecundity*: “there is no fine-tuning, because intelligent life of some form will emerge under extremely varied circumstances” (Tegmark et al. 2006, 4).

We thus need to develop methods, concepts and simulation tools to explore the space of possible universes (the “cosmic landscape" as Leonard Susskin (2006) calls it in the framework of string theory). In (Vidal 2008), I proposed to call this new field of research “Artificial Cosmogenesis”. It is an attempt to a “general cosmology”, in analogy with Artificial Life which appeared with the help of computer simulations to enquiry about a “general biology”.

In summary, if we assume that initial conditions are analogous to incompressible information, then there are two possible reactions. Either we claim that we reached the limit of scientific understanding; or we recognize that we need an extended framework. Multiverse and computer simulations of other possible universes are examples of such extended frameworks.

Let us now see some limits of this computational analogy. If we apply our careful analogical reasoning, we can ask "what is disanalogous between the functioning of our universe and that of a computer?". We can at least make the following restrictions. In a computational paradigm, space and time are assumed to be independent, and non-relativistic. Most of the well studied cellular automata even use only two spatial dimensions, which is of course a limitation for complexity to develop.

A fundamental difference between a physical and an informational-computational paradigm is that the former has at its core *conservation laws* such as the conservation of energy, where the total amount of energy remains unchanged in the various transformations of the system. By contrast, the bits manipulated by computers are not subjected to such conservation laws. We neither create nor destroy energy, whereas we easily create and delete files in our computers. This difference can be summarized by the adage: "When sharing energy, we divide it. When sharing information, we multiply it." (formula that I borrow from Pierre-Alain Cotnoir).

Another limit of this computational paradigm (which is similar to a Newtonian worldview) is that when we have initial conditions, and a set of rules or laws, then the evolution of the system is trivial and predictable: it is just an application of rules/laws to the initial conditions. There would not be much more to understand, as such a model would capture the whole of reality.

The complexity of interactions (such as synergies, feed-back loops, chaos, random errors, etc.) is not in the focus of this approach. The biological analogy is more appropriate in giving insights into the complexity of interactions. Embryologists know that the formation of a fetus is a process of an incredible and fascinating complexity, leading from one single cell, to the complexity of a billions-cells organism. The development of the individual is certainly not as easy to predict from the genome to the phenotype as was the case with the computational paradigm: we just needed the initial conditions and a set of rules to understand the dynamic. By contrast, in biology, phenomena of phenotypic plasticity have been identified, i.e. the acknowledgement that phenotypes are not uniquely determined by their genotype. This becomes particularly clear when considering genetically identical twins. They exhibit many identical features, but also a unique differentiation, due to more stochastic processes occurring during the development. As Martin Rees (2000, 21) noticed, cosmology deals with the inanimate world, which is in fact simpler than the realm of biology. A phenomenon is difficult to understand because it is complex, not because it has a huge extension.

# 4 The biological universe

The idea that our universe is similar to an organism has a long story, which is still very inspiring. It can be traced back to the Ancient Greece (see Barrow and Tipler 1986 for historical aspects). One general aim of the "Evo Devo Universe" research community is to explore how traditional cosmology can be enriched by introducing a biological paradigm, as suggested by George Ellis (2007b, Thesis H4). More specifically, the field of evolutionary developmental ("evo-devo") biology (e.g. Carroll 2005) provides a great source of inspiration, acknowledging both the contingency of evolutionary processes and the statistically predictable aspect of developmental processes. We will now focus our attention on what is analogous in biological systems regarding initial conditions.

Lee Smolin's Cosmological Natural Selection (CNS) hypothesis is an example of a biologically-inspired approach to the fine-tuning problem (Smolin 1992; 1997; 2007). One can introduce his theory with an (other!) analogy. The situation in contemporary cosmology is analogous to the one in biology before the theory of evolution, when one of the core questions was "*(1) Why are the different species as they are?"*. It was assumed more or less implicitly that "*(2) Species are timeless categories"*. In present physics, the question behind the fine-tuning problem is "*(1') Why are the physical constants as they are?"*. Currently, it is usually assumed (probably from the remains of the Newtonian worldview) that "*(2') Physical constants are timeless"*. It is by breaking assumption (2) that Darwin was able to theorize about the origin of species. Analogously, Smolin is trying to break assumption (2'), by theorizing about the origin of constants.

According to this natural selection of universes theory, black holes give birth to new universes by producing the equivalent of a big-bang, which produces a baby universe with slightly different constants. This introduces variation, while the differential success in self-reproduction of universes (via their black holes) provides the equivalent of natural selection. This leads to a Darwinian evolution of universes, whose constants are fine-tuned for black hole generation, a prediction that can in principle be verified.

One might be surprised by the speculative aspect of this theory. Although Smolin emphasizes the refutability of CNS and thus its scientific aspect in (Smolin 2007), he himself is not proud that the theory talks about processes outside the universe (Smolin 1997, 114). This conjectural aspect of the theory puts it at the edge of science and philosophy (Vaas 1998). Let us stress once again that when we attempt to answer the metaphysical question "why is the universe the way it is?", we must be ready to cross the border of current experimental and observational science. Attempting to answer this issue leads to the construction of *speculative* theories. The nature of the inquiry then becomes philosophical, because we aim at answering our most profound questions here and now, whatever their difficulty and our limited knowledge (Vidal 2007).

This non-scientific (but not un-scientific) aspect would at first sight be a reason to fall into an intellectual relativism, where every theory would have equal value, since anyway, there seems to be no way to *scientifically* compare competing speculations. This is correct, but it is still possible to compare speculations from a *philosophical* standpoint. I proposed a general philosophical framework and criteria to compare philosophical speculations (Vidal 2007; 2009). In the following pages, we will intentionally address the problem from this speculative philosophical point of view, focusing our attention on the *objective criteria* we have identified for a good philosophical system. A philosophical system is better than an other, when, other things being equal:

(1) It has a better fit with all the natural sciences.

(2) It addresses and adequately resolves a broader range of philosophical questions.

(3) It exhibits greater internal and systemic coherence. It thus has fewer anomalies.

From this point of view, CNS has some limitations. Firstly, the roles of life and intelligence are incidental. Criterium (2) is then poorly satisfied because, without including life in the cosmic picture, this theory diminishes its philosophical scope and depth considerably. Indeed, the philosophical endeavour is mainly concerned with the relation between humanity and intelligence on the one hand and the universe on the other hand. Secondly, the theory does not propose a hereditary mechanism for universe replication (Gardner 2003, 84). Its internal coherence is thus imperfect and could be improved regarding criteria (3). As Gentner and Jeziorski proposed in the third principle of analogical reasoning (see the second section), one should seek for the greatest possible *systematicity* in the relational network. Is it possible to overcome these two shortcomings?

A few authors have dared to extend CNS by including intelligent life into the picture (Crane 2009; Harrison 1995; Gardner 2000; 2003; Baláz 2005; Smart 2008; Vidal 2008; Stewart 2009). They put forward the hypothesis that life and intelligence could perform this mechanism of heredity, thus playing an essential role in the Darwinian evolution of universes. To better grasp this extension of CNS, Baláz and Gardner proposed to consider von Neumann's (1951) four components of a self-reproducing automaton. We summarized this completion of CNS in table 1. below.

Let us describe these four components in more detail. Physical constants are analogous to DNA in biology, and to the *blueprint* of this self-reproducing automaton. The universe at large or the cell as a whole constitute the *factory.* When furnished with the description of another automaton in the form of a blueprint, the factory will construct this automaton. The *reproducer* reads the blueprint and produces a second blueprint. In the cell, these are reproduction mechanisms of the DNA. The *controller* will cause the reproducer to make a new blueprint, and cause the factory to produce the automaton described by the new blueprint. The controller separates the new construction from the factory, the reproducer and the controller. If this new construction is given a blueprint, it finally forms a new independent self-reproducing automaton.

| ***Components*** | ***Description*** | ***BIOLOGY (cell)*** | ***COSMOLOGY (universe)*** |
|---|---|---|---|
| Blueprint | Gives instructions for the construction of the automaton | Information contained in the DNA | Physical constants |
| Factory | Carries out the construction | Cell | The universe at large |
| Reproducer | Reads the blueprint and produces a second blueprint | The reproduction of the DNA | CNS:? **Intelligence unravelling the universe's blueprint** |
| Controller | Ensures the factory follows the blueprint | The regulatory mechanisms of the mitosis | CNS:? **A cosmic process, aiming at universe reproduction** |

Table 1. Components of a von Neumann's (1951) self-reproducing automaton. The second column provides a general description of the automaton functions. The third and fourth columns propose examples respectively in biology - the cell - and in cosmology - the universe.

We now clearly see the limits of the CNS model, which is not specifying what the reproducer and controller are. Intelligence unravelling the universe's blueprint can precisely fulfil the reproducer's function. This reproducer component is indeed essential for providing a mechanism for heredity. Without heredity, there can be no Darwinian evolution. The controller in this context would be a more general process, aiming at universe reproduction with the help of intelligence. In Table 1, we completed in bold these two missing components of CNS, thus including intelligence in this hypothesized cosmic reproduction process.

Let us now explore in more depth the biological analogy of natural selection and apply it to cosmological evolution. Natural selection implies a differential reproduction with a high level of fidelity, still permitting some mutations. Is there a fidelity in reproduction of physical constants in CNS? If the fidelity were perfect, there would be no evolution, and thus no dynamic to fine tune the constants. If the fidelity were not perfect, it would imply that there are slight changes in the constants from one universe to another. Yet the whole point of fine-tuning arguments is to show that small changes in physical parameters do not lead to a viable universe! It thus seems very unlikely that a CNS would succeed in producing a fine-tuned universe.

A consequence of this speculative theory is that intelligent life, unravelling the universe through scientific understanding, generates a "cosmic blueprint" (a term used by Paul Davies (1989)). The cosmic blueprint can be seen as the set of physical constants; or just initial conditions of a cosmological model, if our previous reasoning holds. One can now throw a new light on the fact that cosmic evolution gave rise to scientific activity. In this view, the increasing modelling abilities of intelligent beings is not an accident, but an indispensable feature of our universe, to produce a new offspring universe.

I have argued that fine-tuning of this cosmic blueprint would take place in "virtual universes", that is in simulated universes (Vidal 2008). This scenario is likely if we seriously consider and extrapolate the exponential increase of computer resources. As argued earlier, simulating universes would also allow virtual experimentation, and thus to compare our own universe with other possible ones. This endeavour would therefore help to progress concretely on the fine-tuning issue.

One can interpret this approach as a variation on the multiverse proposal. However, the selection of universes would take place on *simulated universes*, replacing Smolin's natural selection of *real universes* (Barrow 2001, 151). In CNS, we need many generations of universes in order to randomly generate an interesting fine-tuned universe. In contrast, these simulations would dramatically improve the process by artificially selecting (via simulations) which universe would exhibit the desired features for the next generation universe. This would facilitate the daunting task of producing a new universe. In this case it would be appropriate to speak about a "Cosmological *Artificial* Selection" (CAS), instead of a "Cosmological *Natural* Selection".

One question which naturally arises when considering this CAS, is "why would intelligent life want to produce a new offspring universe?". This is an issue which would deserve much more development, but we briefly propose the following answer (see also (Harrison 1995; Vidal 2008)). The core problem an intelligent civilization has to deal with in the very far future is the inevitable thermodynamical decay of stars, solar systems, galaxies and finally of the universe itself. This scenario is commonly known as the *heat death* (e.g. Adams and Laughlin 1997; Ćirković 2003 for a review). Surprisingly there is little interest in this fundamental and inevitable ageing of the universe. In the realm of biology, the solution to the problem of aging is reproduction. Could it be that an analogous solution would take place at the scale of the universe? Pursuing this cosmic replication process would in principle enable the avoidance of heat death in a particular universe (Vidal 2008). Cosmic evolution would then continue its course indefinitely.

This approach may sound like science-fiction. It is however no more extravagant than the competing explanations, which are the belief in a supernatural God, or the ontological statement that there actually exist a huge number or an infinity of other universes. To summarize, the perspective of a CAS:

1. Proposes a response to the heat death problem.
2. Enables us to scientifically progress on the fine-tuning problem, via universe simulations. This remains acceptable even if one does not wish to endorse the complete speculative framework proposed, i.e. with intelligent life involved in universe reproduction.
3. Offers intelligent life a fundamental role in cosmic evolution.

# 5 Conclusion

Our analysis of physical parameters led us to the conclusion that fine-tuning of physical constants would progressively be reduced to the fine-tuning of initial conditions in a cosmological model. However, it is unlikely that a physical theory would derive *all* cosmological initial conditions. Multiverse and universe simulations are two similar options to go beyond this limitation, because they both propose ways to vary those initial conditions.

The computational analogy suggests a description of the universe in terms of information and computation. Simulations have already started to be used to tackle the fine-tuning problem, allowing us to vary *several* physical parameters, and exploring the resulting possible universes.

We outlined a *philosophical* extension of Smolin's "Cosmological Natural Selection" theory, inspired by a biological analogy. It led us to the hypothesis that intelligent life is currently unravelling a cosmic blueprint, which could be used in the far future to produce a new offspring universe. Simulating an entire universe would be an efficient way to fine tune this cosmic blueprint. Furthermore, this offspring universe would perpetuate cosmic evolution beyond the predictable heat death of our universe.

Physical, computational and biological descriptions are different ways to model the universe. Providing that we make clear and explicit our analogical reasoning, keeping this diversity of possible modelling approaches is certainly a way to constantly enrich and cross-fertilize our models. Importantly, we have to acknowledge the limits of scientific enquiry when tackling the fine-tuning problem. A philosophical approach allows us to keep a rational and critical attitude.

## 6 Acknowledgements

I thank Sean Devine, Börje Ekstig, Pushkar Ganesh Vaidya, Bernard Goossens, Francis Heylighen, Gerard Jagers op Akkerhuis, John Leslie, Robert Oldershaw, Marko Rodriguez and John Smart for useful and stimulating comments. Thanks to Sophie Heinrich and Jean Chaline for insightful criticisms about the biological analogy in cosmology. I warmly thank Luke Lloyd for his help in improving the English.